\documentclass[10pt,letterpaper]{article}
\usepackage{opex3}
\usepackage[]{graphicx}

\newcommand{\figwidth}{0.775}
\begin{document} 

\title{Maximizing the mode instability threshold of a fiber amplifier} 

\author{Arlee V. Smith$^*$ and Jesse J. Smith}

\address{AS-Photonics, LLC, 8500 Menaul Blvd. NE, Suite B335, Albuquerque, NM USA 87112}

\email{$^*$arlee.smith@as-photonics.com}


\begin{abstract}
We show by detailed numerical modeling that stimulated thermal Rayleigh scattering can account for the modal instability observed in high power fiber amplifiers. Our model illustrates how the instability threshold power can be maximized by eliminating amplitude and phase modulation of the signal seed and the pump, and by careful launch of the signal seed. We also illustrate the influence of photodarkening and mode specific loss on the threshold.
\end{abstract}



\section{INTRODUCTION}\label{sec:intro}

\begin{figure}[h]
\centering
\includegraphics[width=\figwidth\textwidth]{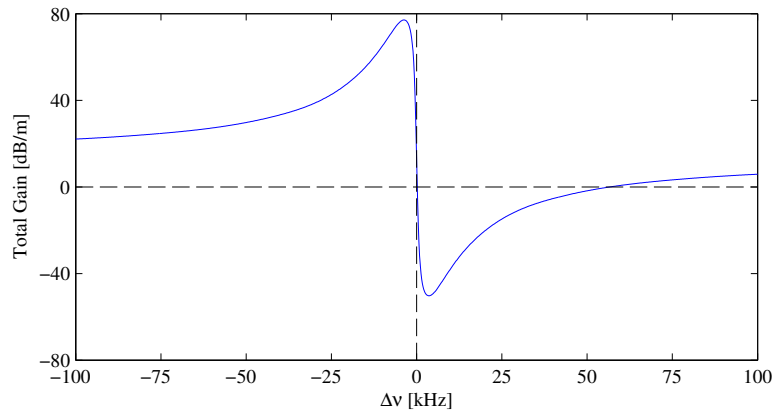}
\caption{\label{fig:gain_vs_freq}Sample gain of mode LP$_{11}$ versus its frequency offset from mode LP$_{01}$. The gain includes both laser gain and mode coupling gain due to STRS. Laser gain accounts for the upward shift of approximately 15 dB/m.}
\end{figure}

Above a sharp power threshold the output beam from a multimode fiber amplifier is radically altered\cite{Eidam2011,Otto2012,Haarlammert2012,Stutzki2011a}. If the fundamental LP$_{01}$ mode is injected at the fiber input, slightly above the threshold the output is largely in LP$_{11}$. This modal instability is caused by a stimulated thermal Rayleigh scattering (STRS) process\cite{Smith2011,Smith2012,Hansen2012}. Quantum defect heating creates a temperature grating, and consequently a refractive index grating, that is responsible for coupling between the two fiber modes. Numerical models of the STRS process show that, like other stimulated Rayleigh scattering processes, the gain has a dispersive shape\cite{Smith2012,Hansen2012} similar to that shown in Fig.~\ref{fig:gain_vs_freq}. A frequency offset between the two coupled modes is necessary to produce the phase shift essential for mode coupling\cite{Smith2011,Hansen2012}, and power transfer is toward the mode with the lower frequency.

The purpose of this paper is to provide some guidance on how the mode instability threshold can be held at the highest possible power. There are two approaches to this: optimizing the fiber design, or optimizing the operating conditions. This paper is primarily concerned with finding the optimum operating conditions.

\section{NUMERICAL MODEL}\label{sec:numerical}

\begin{figure}[h]
\centering
\includegraphics[width=\figwidth\textwidth]{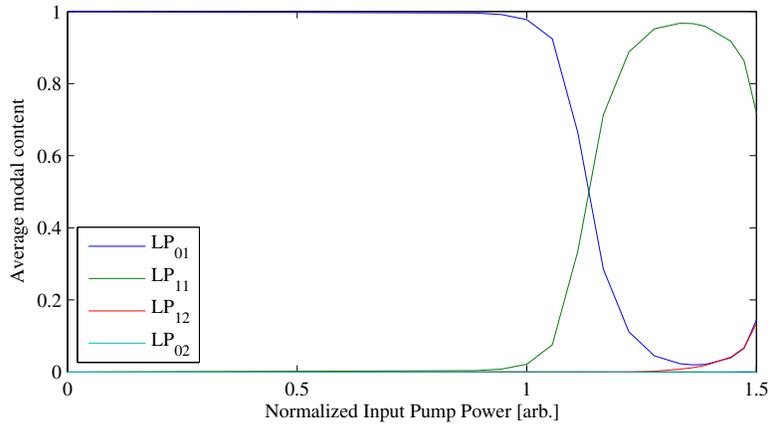}
\caption{\label{fig:modal_contents}Fraction of the signal power in various modes versus the input pump power (in normalized units). Here, the frequency shifted LP$_{11}$ is seeded by quantum noise simulated by $10^{-16}$ W. The sharp mode instability threshold occurs near the normalized pump power of 1.0.}\end{figure}

We reported elsewhere our detailed numerical model of the STRS process\cite{Smith2011,Smith2013a} based on the assumption that the powers of the two coupled modes and the temperature vary only periodically. This steady-periodic assumption allows the use of a steady-periodic Green's function to compute the time-dependent temperature profiles responsible for mode coupling\cite{Cole2006}. The Green's function temperature solver is combined with a split-step fast-Fourier transform beam propagation method to model mode coupling. This approach has the advantage that all fiber modes are simultaneously included, and the population inversion profiles are realistic. The highly numerical nature of the model makes it relatively easy to add other physical effects. This model allows us to predict the frequency shift between LP$_{01}$ and LP$_{11}$ which produces maximum mode coupling gain. Our predicted frequencies are in good agreement with observed signal output modulation frequencies\cite{Otto2012,Ward2012}. Another key feature of the model is the instability threshold is well-defined, with a sharp change of LP$_{11}$ content with increasing pump power, as illustrated in Fig.~\ref{fig:modal_contents}.

In the following sections we will examine how the instability threshold is impacted by pump and signal amplitude modulation, photodarkening, and mode specific loss. We also suggest the possibility of restoring the threshold by countering the pump modulation with signal modulation. As a baseline case, without any of these influences, we model the LPF45 fiber amplifier described by Otto {\it et al.}\cite{Otto2012}, using the parameters listed in Table \ref{tab:test}. This is a photonic crystal fiber, but we simulate it using a co-pumped, step index fiber with the core size and numerical aperture adjusted to give an LP$_{01}$ mode close to the reported size. For the baseline amplifier, we use a signal seed power of $10^{-16}$ W in the frequency shifted LP$_{11}$ mode to simulate quantum noise seeding. This is calculated using a noise spectral power density of $h\nu$ multiplied by the amplified bandwidth of approximately 500 Hz. The baseline amplifier has a threshold at a normalized pump input power of 1.0. We will compare other thresholds to this normalized value. Threshold is defined here as the pump input power at which 5\% of the output signal is in the higher order mode LP$_{11}$. Since this paper is only concerned with qualitative behavior, and we expect similar behaviors for co- and counter-pumped fibers, we model only the co-pumped case. The experimental mode coupling performance of the LPF45 fiber has been reported\cite{Otto2012}, permitting comparisons of some of our predictions with laboratory results.

\begin{table}[tbh]
\caption{LPF45 fiber amplifier parameters used in the numerical model.} 
\label{tab:test}
\begin{center}       
\begin{tabular}{|ll|ll|}
\hline
\rule[-1ex]{0pt}{3.5ex} $d_{\rm core}$ & 81 $\mu$m & $d_{\rm dope}$ & 63 $\mu$m  \\
\rule[-1ex]{0pt}{3.5ex} $d_{\rm clad}$ & 255 $\mu$m & $P_{\rm 01}$ & 10 W   \\
\rule[-1ex]{0pt}{3.5ex}  $\lambda_{\rm pump}$ & 976 nm & $\lambda_{\rm signal}$ & 1060 nm \\
\rule[-1ex]{0pt}{3.5ex} $\sigma^a_{\rm pump}$ & $2.47\times 10^{-24}$ m$^2$ & $\sigma^e_{\rm pump}$ & $2.44\times 10^{-24}$ m$^2$  \\
\rule[-1ex]{0pt}{3.5ex} $\sigma^a_{\rm sig}$ & $5.8\times 10^{-27}$ m$^2$ & $\sigma^e_{\rm sig}$ & $2.71\times 10^{-25}$ m$^2$  \\
\rule[-1ex]{0pt}{3.5ex} $n_{\rm core}$ & 1.45015 & $n_{\rm clad}$ &1.45\\
\rule[-1ex]{0pt}{3.5ex} $N_{\rm Yb}$ & $3.0\times 10^{25}$ m$^{-3}$ & $L$ & 1.2 m\\
\hline 
\end{tabular}
\end{center}
\end{table} 

\section{MODULATED SIGNAL SEED}\label{sec:modulatedseed}

While seeding the LP$_{11}$ mode with quantum noise is unavoidable, additional sources of frequency shifted LP$_{11}$ light can dramatically reduce the threshold. For example, if the input signal light is amplitude modulated, frequency components within the STRS amplification band can be populated. If some of this frequency shifted seed light is accidentally injected into mode LP$_{11}$, it will seed the amplified mode. Among other possibilities, such amplitude modulation will be present if the seed is an ASE source. It is also present at some level in any other source of the seed light.

\begin{figure}[h]
\centering
\includegraphics[width=\figwidth\textwidth]{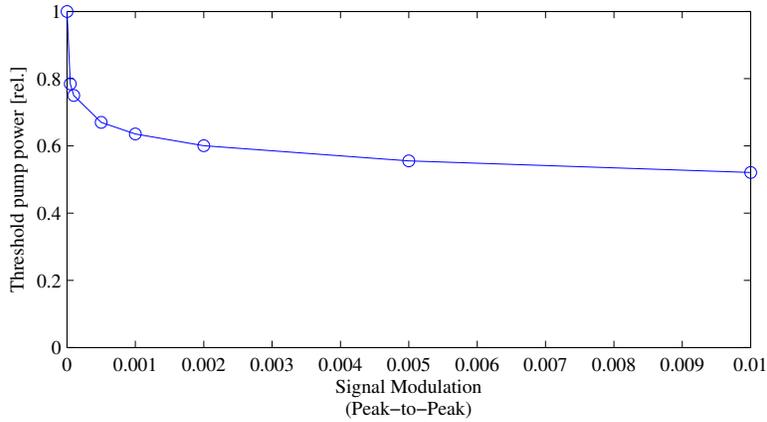}
\caption{\label{fig:threshold_vs_signalAM}Normalized pump power at mode instability threshold vs. magnitude of signal amplitude modulation. The magnitude is defined as the peak-to-peak power variation normalized to the average power. Threshold is defined as 5\% of the signal output in LP$_{11}$. The leftmost point corresponds to the baseline case.}
\end{figure}

We model an amplitude modulated seed with 9.9 W injected into LP$_{01}$ and 0.1 W injected into LP$_{11}$. The frequency is adjusted for maximum mode coupling gain. Figure~\ref{fig:threshold_vs_signalAM} demonstrates the influence of increasing depth of signal modulation on the threshold where the baseline case corresponds to zero modulation. As the figure shows, the threshold is strongly reduced even for small levels of modulation.

The important thing is the strength of the seed in the frequency-shifted light in LP$_{11}$. The threshold reduction is similar if only the light in LP$_{11}$ is amplitude modulated rather then all the seed light as was the case for the results presented in Fig. \ref{fig:threshold_vs_signalAM}. This means small mechanical vibrations that affect the amount of light accidentally injected into LP$_{11}$ could produce the amplitude modulation in the amplified band. It is also unimportant whether the seed light is amplitude modulated or phase modulated. They produce similar threshold reductions.

\section{MODULATED PUMP}\label{sec:modulatedpump}

\begin{figure}[h]
\centering
\includegraphics[width=\figwidth\textwidth]{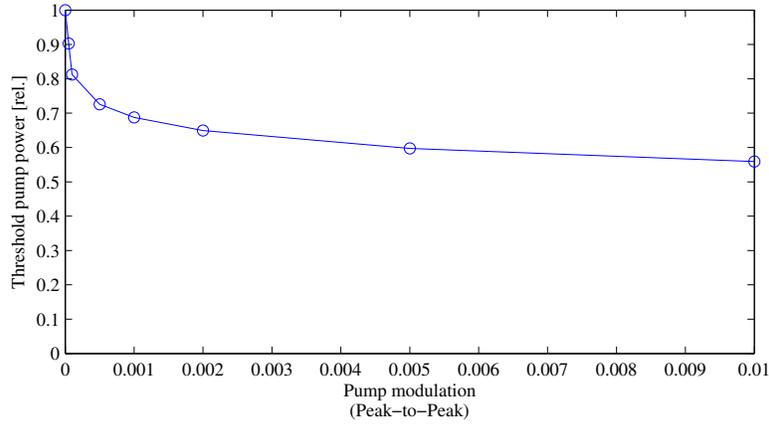}
\caption{\label{fig:threshold_vs_pumpAM}Normalized pump power at mode instability threshold vs. magnitude of pump amplitude modulation. The magnitude is defined as the peak-to-peak power variation normalized to the average power. Threshold is defined as 5\% of the signal output in LP$_{11}$.}
\end{figure}

Even if the signal seed is unmodulated, pump modulation produces almost the same effect as signal modulation. A modulated pump quickly impresses amplitude modulation on the signal light in LP$_{11}$, leading to population of the frequency shifted signal sideband. We model this effect by seeding LP$_{01}$ with 9.9 W and LP$_{11}$ with 0.1 W, both unmodulated. We modulate the input pump by varying amounts at the frequency of maximum mode coupling gain. The results are shown in Fig.~\ref{fig:threshold_vs_pumpAM} with the baseline case corresponding to zero modulation. Again a small amount of modulation leads to a strong reduction in threshold.

\section{COUNTER MODULATION OF SEED AND PUMP}\label{sec:countermodulation}

\begin{figure}[h]
\centering
\includegraphics[width=\figwidth\textwidth]{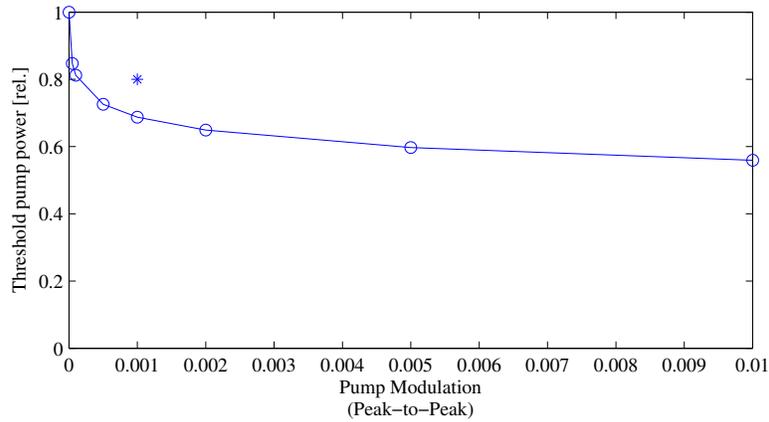}
\caption{\label{fig:threshold_vs_pumpAM_countered}This is the same as Fig.~\ref{fig:threshold_vs_pumpAM} with the addition of the star indicating the improvement by optimized counter modulation of the input signal.}
\end{figure}

The similarity of Figs.~\ref{fig:threshold_vs_signalAM} and \ref{fig:threshold_vs_pumpAM} suggests that perhaps a small pump modulation can be counteracted by an appropriately adjusted signal modulation. We tested this idea using a pump amplitude modulation of 0.001, and show in Fig.~\ref{fig:threshold_vs_pumpAM_countered} the results when the signal modulation is optimized in amplitude and phase. We could not fully restore the threshold, but did improve it by 15\%.

\section{PHOTODARKENING}\label{sec:photodarkening}

Photodarkening can also strongly reduce the instability threshold. We model this using a very simple photodarkening model. We assume a uniform linear absorption of the signal light across the full doped region of the fiber. A more realistic model would account for the transverse shape of the absorption at each $z$ location, but our model should give a reasonable indication of the influence of photodarkening. It also shows the effect other sources of linear absorption might have. We assume the absorbed power is fully and instantaneously converted to heat, so the heating profile matches the irradiance profile over the doped region.

\begin{figure}[h]
\centering
\includegraphics[width=\figwidth\textwidth]{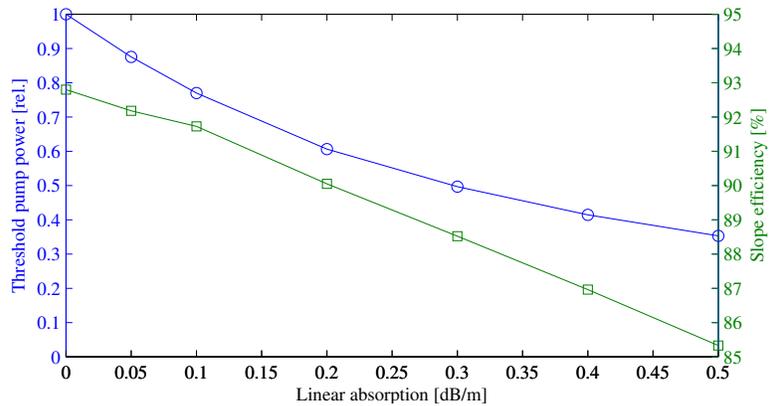}
\caption{\label{fig:threshold_and_efficiency}Threshold (circles)  and efficiency (squares) dependence on linear absorption. Here, LP$_{01}$ is seeded with 10 W and LP$_{11}$ is seeded with $10^{-16}$W at the frequency of maximum gain. Efficiency is defined as the increase in signal power divided by the reduction in pump power.}
\end{figure}

Figure~\ref{fig:threshold_and_efficiency} shows in the blue trace (circles) how the threshold falls with increasing signal absorption. It is important to realize that at the absorptions considered here the efficiency of the amplifier is only slightly reduced, as shown in the green curve (squares). A reduction of 60\% in threshold corresponds to a reduction in efficiency of only 7\%. One important signature of actual photodarkening is that it turns on gradually in the presence of pumping, and it can be reversed by optical bleaching or thermal bleaching\cite{Laurila2012a,Laurila2012b}.

\section{MODE SPECIFIC LOSS}\label{sec:modeloss}

\begin{figure}[h]
\centering
\includegraphics[width=\figwidth\textwidth]{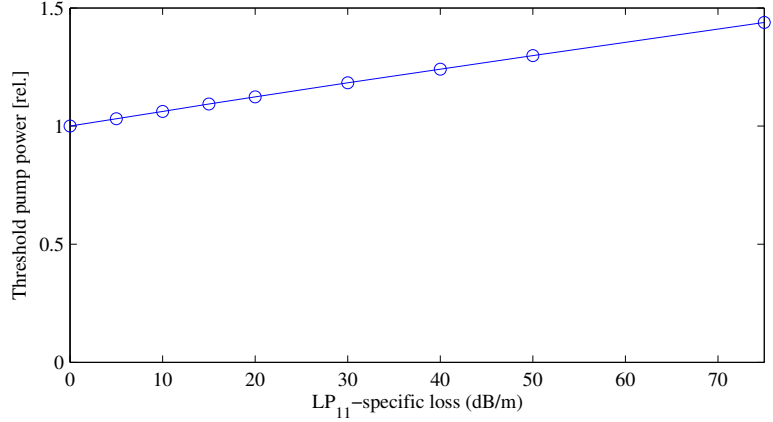}
\caption{\label{fig:threshold_vs_LP11_loss}Influence of mode LP$_{11}$ loss on the relative threshold pump power in a quantum seeded amplifier.}
\end{figure}

If the loss of LP$_{11}$ can be made large without significantly affecting LP$_{01}$, it should be possible to increase the instability threshold while maintaining overall efficiency. Such loss for LP$_{11}$ might be created by bend loss or by clever fiber engineering. We introduce an LP$_{11}$ loss that is constant along the full length of the fiber, with zero loss to LP$_{01}$, with the results shown in Fig.~\ref{fig:threshold_vs_LP11_loss}. We seed LP$_{01}$ with 10 W and the frequency shifted LP$_{11}$ with $10^{-16}$ W so the zero loss point is again the baseline case. It is clear from the figure that large mode specific losses are required to significantly increase the threshold. This is not surprising because the baseline mode coupling gain is larger than 170 dB at threshold, and the gain is approximately linear in pump power.

Several studies of mode discrimination by fiber design or bending have suggested the possiblity of large mode specific loss\cite{Uranus2004,Jiang2006,Ward2008,Stutzki2011b}. Unfortunately, it is common that reports of experimental observations of mode instability thresholds neglect to include mention of mode specific losses. 

\section{CONCLUSION}\label{sec:conclusion}

Careful characterization of fiber amplifiers will be required to maximize the mode instability threshold. All pump lasers and seed sources have some degree of spectral and amplitude modulation, and we showed that even small modulations can strongly reduce the threshold. It should be clear that the pump and seed inputs must be extremely well controlled to maximize thresholds, and that their characterization is an essential part of meaningful reports on modal instability. Additionally, precise accounting of the signal and pump powers may be necessary in order to detect any small linear losses since they also strongly reduce thresholds. Further, it is clear that attempts to raise the mode instability threshold significantly by engineering losses for the higher order modes will require large losses. Doubling the threshold from the baseline requires about 170 dB of loss for LP$_{11}$.

\end{document}